# An Analysis of Speculative Window Decoders for Quantum Error Correction

Jocelyn Li
Princeton University
jl1543@princeton.edu

Margaret Martonosi
Princeton University
mrm@cs.princeton.edu

## I. Introduction

Quantum computing promises significant speedup for classically intractable problems, but realizing this potential requires fault-tolerant quantum computing (FTQC) that uses quantum error correction (QEC) to protect against noise. QEC employs a classical decoder, which analyzes QEC syndrome measurements to track and correct errors during execution. The decoder must operate in real time alongside the QPU, because of blocking operations, which require the current error state to be known before execution can proceed. Consequently, decoder speed can limit overall system performance.

Speculative window decoding [1] can improve decoder performance by partitioning the decoding graph into multiple windows and speculating on dependencies between windows. Partitioning improves performance because the decoder has less data to act on [2]. However, windows must wait for prior dependent windows to finish decoding before they can begin. By speculating on inter-window dependencies, wait time is reduced, as windows can begin decoding before predecessors finish. These speculated dependencies are later verified by a verifying decoder, and a window is restarted if its speculated dependencies were incorrect. Speculative decoding increases decoder throughput and decreases the latency between a window's generation and its successful decoding. Improving this classical-QPU interaction improves overall performance.

However, speculative decoders have only been evaluated under the regime of superconducting qubits, which have fast gate speeds [3]. Different quantum technologies, such as trapped ions and neutral atoms, have different gate speeds that are typically slower [4], [5]. Therefore, we evaluate speculative decoder performance under slow gate speeds to examine its potential utility for different quantum technologies.

Furthermore, speculative decoders have only been evaluated for surface codes and matching decoders [6]–[8]. However, factors such as speculation accuracy, decoder latency, and workload parallelism may vary across different QEC codes, decoders, and hardware platforms [9], [10]. Prior work [1] also has not examined the performance impacts of having fewer decoder processors than the number of windows that are ready to be decoded, which could occur due to a high speculative decoding workload. Here, we examine how these factors impact performance and derive design principles accordingly.

Finally, prior work suggests that speculative decoding will always outperform non-speculative decoding because it will never increase dependency wait times. However, we investigate whether different factors can decrease speculative decoder performance enough for non-speculative decoders to outperform it. This can inform whether speculative decoding will yield significant performance improvements over non-speculative decoding for certain hardware platforms or QEC codes. It can also guide the design of adaptive decoders, which only speculate under conditions that yield higher performance. The main contributions of this work are as follows:

- Under fast gate speeds, speculative decoder performance is more severely bottlenecked by decoder reaction time than under slow gate speeds. Thus, it is more sensitive to changes in speculation accuracy, decoder latency, and processor count. Subsequently, performance under slow gate speeds is higher when reaction time is long.
- Under fast gate speeds, performance is more sensitive to decode-processor count and speculation accuracy than decoder latency. Under slow gate speeds, performance is more sensitive to decoder latency and processor count than speculation accuracy.
- Non-speculative decoding outperforms speculative decoding under fast gate speeds when reaction time is slow.
- To improve performance when the number of ready windows exceeds processor count, we present the *Shallow Speculations First* decoding ordering strategy, which decodes windows in order of increasing speculation depth.
- Under limited processor counts, the most parallel workload may not perform best, since greater parallelism increases the processor count needed for peak performance.

## II. Experimental Methodology

**SWIPER-SIM:** Our experiments use a modified version of SWIPER-SIM [1], a speculative decoder simulator that manages instruction execution, decoding, and speculation. Notably, instruction execution progresses independently of decoder progress until it reaches a blocking operation, which halts execution until its dependent instructions are verified, meaning their corresponding windows are verified by the verifying decoder. To maintain fault tolerance during this wait time, unwanted idle (UI) windows are generated. These windows must be decoded eventually, subsequently increasing the decoder's total workload.

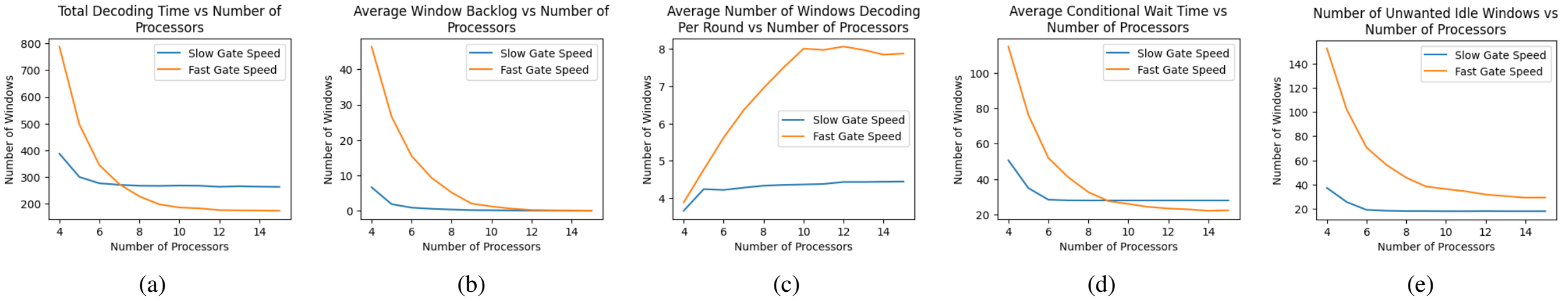

(a) (b) (c) (d) (e)

Fig. 1. Performance metrics as the number of processors varies for fast and slow gate speeds.

**Varied Input Parameters:** SWIPER-SIM takes a lattice surgery schedule, decoder metadata, and speculator metadata as inputs. Here, we vary speculation accuracy, decoder latency, processor count, lattice surgery schedule parallelism, gate speed, and the decoding ordering strategy. We extended SWIPER-SIM to accommodate different gate speeds and ordering strategies.

**Gate Speed:** To model different gate speeds, we use cycle time–the time it takes to perform one round of syndrome measurements–as a proxy for gate speed. Syndrome measurement rounds are obtained by running a syndrome extraction circuit, whose duration depends on gate speed; fast gate speeds correspond to fast cycle times [11].

SWIPER-SIM tracks time at the granularity of superconducting qubit syndrome measurement rounds. Each simulator round corresponds to 1 $\mu$s, according to recent experimental timing [3]. To model slower gate speeds, we introduce the *Slow Gate Multiplier* parameter, which allows each conceptual syndrome measurement round to span multiple simulator rounds. Higher multipliers represent slower gate speeds. By default, our experiments use a fast gate speed with cycle time of 1 $\mu$s and slow gate speed with cycle time of 2 $\mu$s.

**Window Decoding Ordering Strategy:** When the number of windows ready to be decoded exceeds the available processor count, the order in which windows are decoded affects performance. We evaluate three potential orderings: (i) *Shallow Speculations First:* Windows are decoded in order of increasing speculation depth, which is the distance from a window to the most recent generation of windows that are all verified. (ii) *Deep Speculations First:* Windows are decoded in order of decreasing speculation depth. (iii) *Window Generation Order:* Windows are decoded in order of window generation. This is the default strategy used in prior work [1].

**Input Lattice Surgery Schedule:** Our default input lattice surgery schedule contains three T-gate injection circuits operating in parallel on different logical qubits. We use T-gate injection circuits because they commonly occur in Clifford+T circuits and include a blocking Conditional S instruction. This lets us evaluate speculative decoding with a significant performance bottleneck.

To vary workload parallelism, we add/remove T-gate injection circuits on different logical qubits, while normalizing workloads to execute approximately the same total number of injection circuits for fair performance comparison.

## III. Analysis and Results

### A. Performance Bottlenecks for Different Gate Speeds

Figure 1a shows that when processor count is limited, the total decoding time increases more sharply for fast gate speeds than slow gate speeds. This suggests that fast gate speeds are bottlenecked by decoder reaction time, since ready windows wait longer to begin decoding when processors are limited.

Figure 1b also shows that limited processor counts cause a larger window backlog under fast gate speeds than slow gate speeds. A large window backlog suggests that windows are being generated but are waiting to be decoded, demonstrating the decoder bottleneck. In contrast, the slow gate speed's smaller window backlog suggests that the decoder can keep pace relatively well with window generation, indicating a less severe decoder bottleneck.

Figure 1c shows that slow gate speeds have fewer windows being decoded per round than fast gate speeds. This suggests that windows are not being generated quickly enough to support many concurrently decoding windows, indicating a bottleneck in window generation speed. Beyond five processors, processor count exceeds the number of decoding windows for slow gate speeds, suggesting that the decoder is not the bottleneck.

### B. Cascading Effects of Reaction Time for Fast Gate Speeds

Figure 1a shows that the total decoding time increases more sharply for fast gate speeds than slow gate speeds under a limited processor count. Therefore, we hypothesize that the significant performance degradation is primarily caused by the slow decoder reaction time–which results from a limited processor count–and fast window generation speed. Figures 1b, 1d, and 1e suggest that the average window backlog, average conditional wait time, and UI window count explain this performance decline, as their curves follow similar trends.

A slow decoder reaction time causes more windows to accumulate in the backlog because the decoder processes windows more slowly. Subsequently, the conditional wait time before a blocking operation increases because instruction execution must wait for the decoder to finish verifying a larger number of dependent windows that are still in the backlog. This time further increases because the decoder reaction time is slow. More UI windows are subsequently generated during this longer wait time.

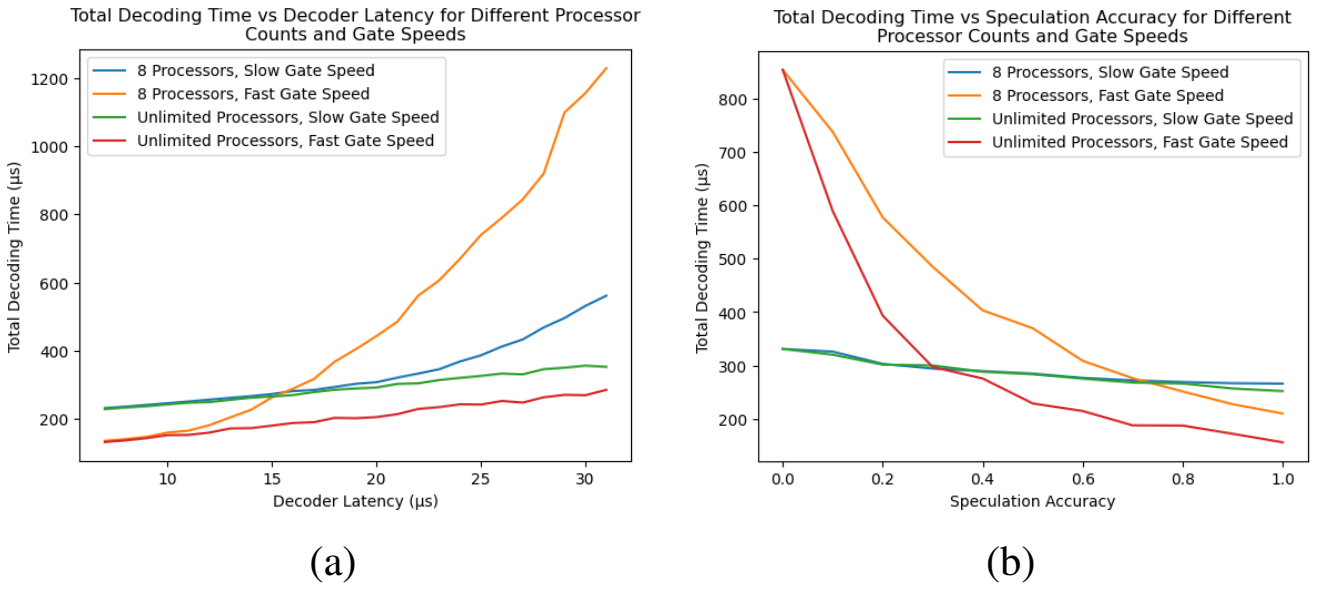

(a) (b)

Fig. 2. Total decoding time under varying decoder latency and speculation accuracy for fast and slow gate speeds with different processor counts.

A fast gate speed also causes the window backlog to increase because windows are generated more quickly than the decoder can process them. By the same logic as above, the conditional wait time and number of UI windows subsequently increases. Notably, the number of UI windows increases more sharply with a fast gate speed, as seen in Figure 1e, since more UI windows are generated within the same time due to their shorter window generation time. Because these two factors independently increase the number of UI windows, the speculative decoder under fast gate speeds has much more extra work, significantly increasing its total decoding time.

### C. Performance Comparison of Fast and Slow Gate Speeds

*1) Decoder Latency:* Figure 2a demonstrates that the performance effects of decoder latency depends on processor count. As processor count decreases, total decoding time becomes more sensitive to increases in decoder latency, suggesting that latency requirements should be more strict when processor count is limited. Performance is more resilient to increases in latency with higher processor counts because windows can begin decoding as soon as their dependencies are calculated, since processor availability is not a constraint. Therefore, though each window takes longer to decode, the overall performance does not decrease significantly, as the decoder can concurrently decode many windows.

This figure also shows that slow gate speeds are more resilient to increases in decoder latency than fast gate speeds, particularly when processor count is limited. Subsequently, slow gate speeds outperform fast gate speeds when latency is high and processor count is limited. This also suggests that latency requirements can be more lenient for slow gate speeds up to a certain threshold, beyond which decoding time increases significantly. This indicates the point at which the performance bottleneck shifts from window generation speed to decoder reaction time.

*2) Speculation Accuracy:* Figure 2b shows that regardless of processor count, slow gate speeds will outperform fast gate speeds if speculation accuracy is sufficiently low. Slow gate speeds produce fewer speculations because their slower window generation results in fewer windows being generated in time for speculation. Therefore, a low speculation accuracy impacts the performance of fast gate speeds more significantly. This implies that speculation accuracy requirements must be stricter for fast gate speeds than slow gate speeds.

*3) Processor Count:* Figures 1a and 2 demonstrate that processor count significantly affects performance, especially for fast gate speeds. This is because fast gate speeds have more windows that can be concurrently decoded, as seen in Figure 1c. Therefore, its requirement for processor count must be stricter. In contrast, slow gate speeds can have more lenient requirements for processor count until a threshold, beyond which performance also significantly declines.

### D. Comparison of Non-Speculative and Speculative Decoding

In this section, we use two slow gate speeds with cycle times of 2 and 3 $\mu$s. We also use a parallel window decoder for the non-speculative decoder.

*1) Fast Gate Speeds:* For fast gate speeds, Figure 3a shows that non-speculative decoding outperforms speculative decoding when processor count or speculation accuracy is sufficiently low. Furthermore, as processor count increases, the speculation accuracy threshold below which non-speculative decoding outperforms speculative decoding decreases.

Speculative decoding's worse performance stems from the cascading performance degradation that arises with a fast gate speed and slow decoder reaction time. This can result from low speculation accuracy or limited processor count. Low speculation accuracy does not affect the non-speculative decoder's reaction time because it does not speculate. Limited processor count also primarily affects speculative decoding here, because non-speculative decoding typically has fewer concurrent decoding windows than speculative decoding, since windows must wait longer for dependencies to resolve. This lowers the processor requirement for non-speculative decoding; its performance does not decrease at processor counts where speculative decoding's performance significantly declines.

*2) Slow Gate Speeds:* In contrast, for slow gate speeds, Figures 3b and 3c show that speculative decoding consistently outperforms non-speculative decoding, regardless of speculation accuracy or the processor count. Moreover, this advantage increases as gate speed decreases.

This is because the slow window generation speed limits the number of UI windows generated during the conditional wait time. Therefore, even if the wait time increases from a slow decoder reaction time, not much extra decoding work is created. Therefore, it is always beneficial to speculate for slow gate speeds–even if performance benefits are smaller–since they do not incur significant penalties from speculating with a slow decoder reaction time.

### E. Different Window Decoding Ordering Strategies

When available processor count is less than the number of windows that are ready to be decoded, the order in which they are decoded affects performance.

*1) Fast Gate Speeds:* For fast gate speeds, Figure 4a shows that *Shallow Speculations First* achieves the highest performance, especially with a limited processor count. This is because decoding windows in order of increasing speculation

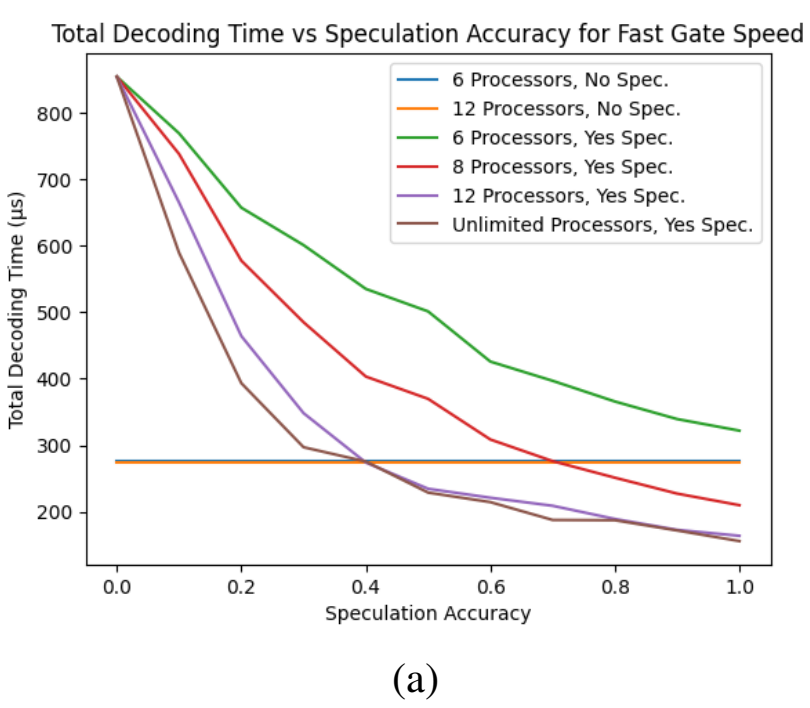

(a)

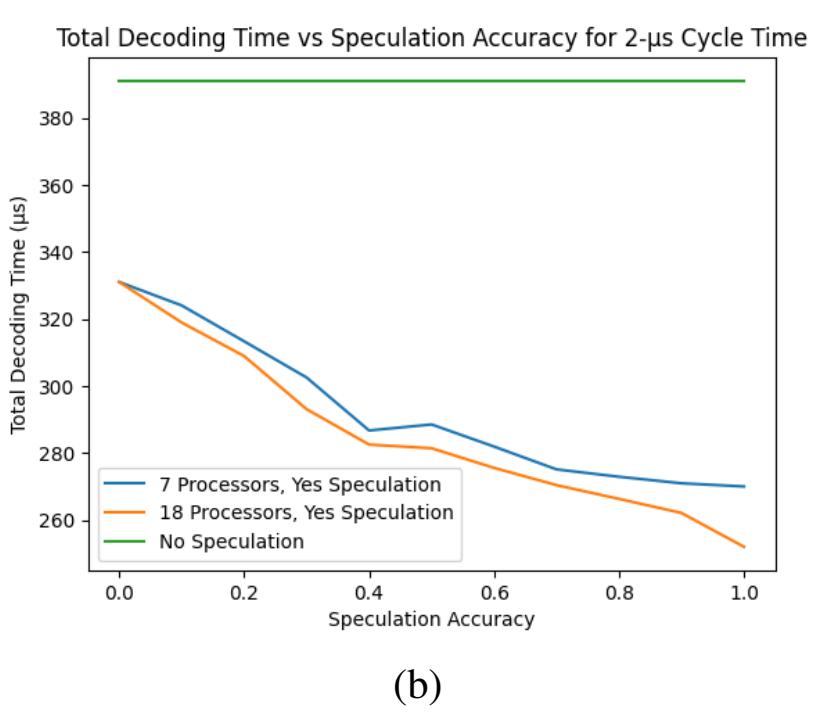

(b)

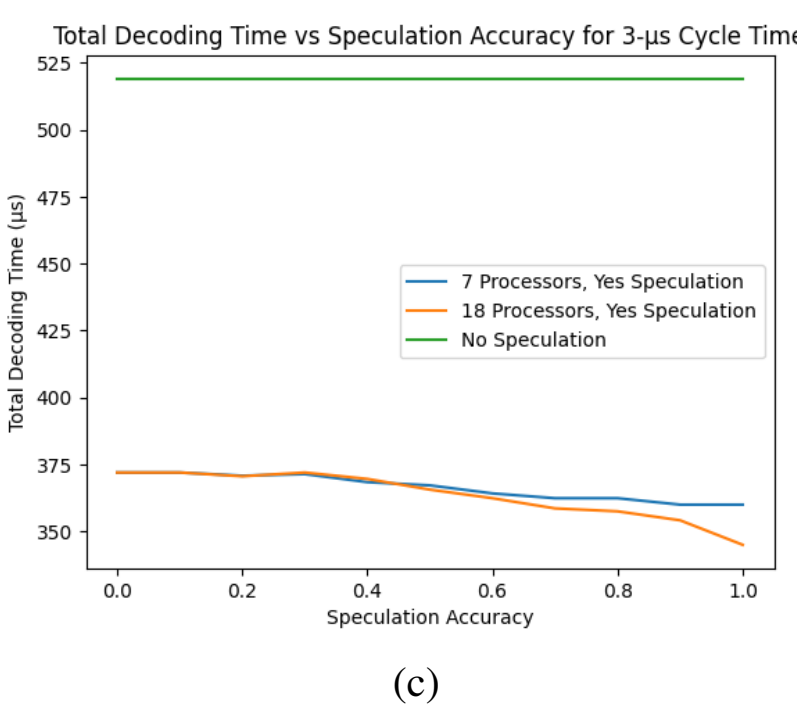

(c)

Fig. 3. Graphs for total decoding time as speculation accuracy varies for different gate speeds. These graphs compare the performance of speculative and non-speculative decoding with different processor counts.

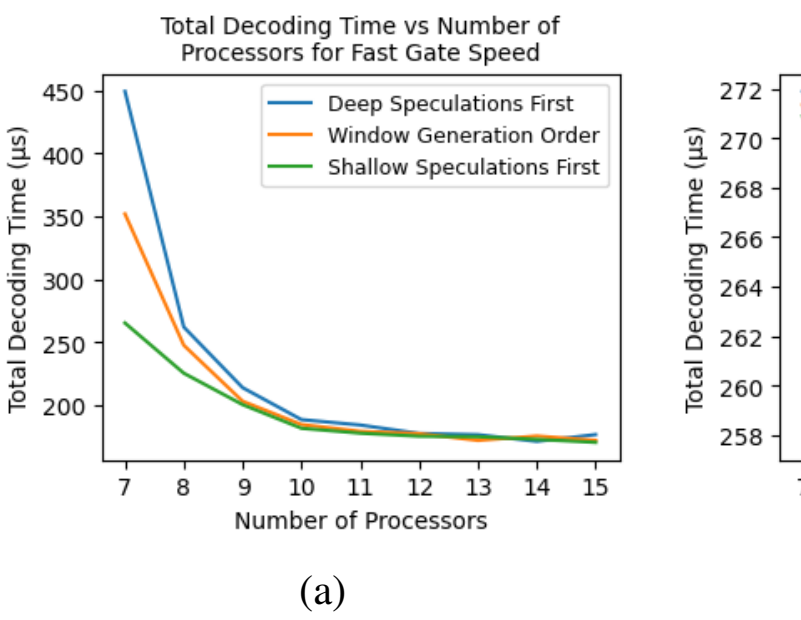

(a)

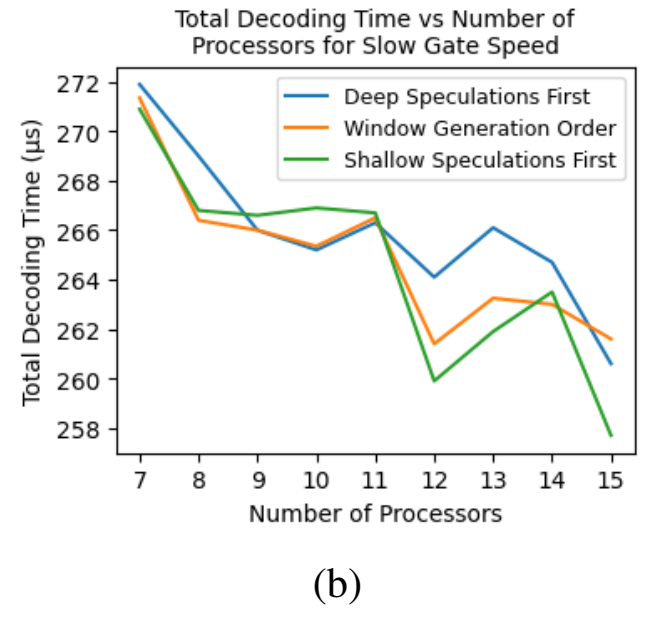

(b)

Fig. 4. Total decoding time as the number of processors varies for different decoding ordering strategies, under fast and slow gate speeds.

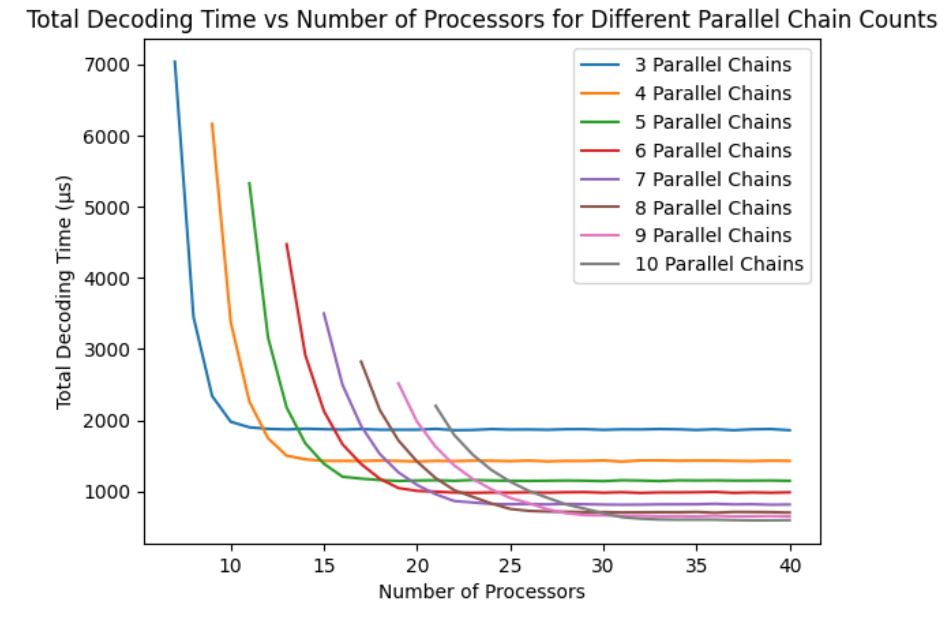

Fig. 5. Total decoding time of workloads with different numbers of parallel chains as processor count varies for fast gate speeds.

depth prioritizes the verifying decoder, since windows that are being verified have speculation depth of one because their predecessors must all be verified. This helps the verifying decoder keep pace with window generation, reducing window backlog, conditional wait time, and number of UI windows, subsequently improving performance.

*2) Slow Gate Speed:* For slow gate speeds, Figure 4b shows that there is no significant performance difference between different ordering strategies. Because of the slow window generation speed, the window backlog is rarely larger than the processor count, causing all strategies to perform similarly.

### *F. Workloads With Differing Amounts of Parallelism*

For fast gate speeds, Figure 5 shows that as workload parallelism increases, the number of processors needed to minimize decoding time also increases because higher parallelism results in more concurrently decoding windows. Furthermore, peak performance improves as parallelism increases.

However, Figure 5 reveals that under limited processor counts, the workload with the highest performance may not have the most parallelism. More parallel chains results in more windows being generated concurrently. Therefore, the verifying decoder requires extra processors to keep pace, as it can concurrently decode windows from different chains. If processor count is limited, the verifying decoder falls behind window generation, increasing conditional wait time and decreasing performance.

## IV. Future Work

One avenue of future work is context-aware confidence predictors. Currently, the speculator always speculates, but circuit context may affect speculation accuracy, particularly if circuit noise varies with context. We find that not speculating outperforms speculating at low speculation accuracies. Therefore, a context-aware confidence predictor could predict per-context speculation accuracy and decide whether to speculate.

Another avenue of future work is migrating speculative decoding to other QEC codes and decoders, using this work's design principles to identify the conditions that would yield the greatest performance improvements.

## V. Conclusion

In this work, we examine the effects of gate speed, speculation accuracy, decoder latency, processor count, workload parallelism, and decoding ordering strategies on speculative decoder performance. From this, we present design principles that can guide developers in identifying the conditions under which speculative decoding yields the most significant performance improvements, along with when non-speculative decoding performs better.